\documentclass[a4paper]{article}

\usepackage{INTERSPEECH2022,enumitem,hyperref,multirow,booktabs}

\title{Transplantation of Conversational Speaking Style with Interjections in Sequence-to-Sequence Speech Synthesis}
\name{Raul Fernandez$^1$, David Haws$^1$, Guy Lorberbom$^1$, Slava Shechtman$^1$, Alexander Sorin$^1$}
\address{ IBM Research AI}
\email {\{fernanra,dhaws\}@us.ibm.com,  guy.lorberbom@ibm.com, \{slava,sorin\}@il.ibm.com}

\begin{document}

\maketitle
\begin{abstract}
Sequence-to-Sequence Text-to-Speech architectures that directly generate low level acoustic features from phonetic sequences are known to produce natural and expressive speech when provided with adequate amounts of training data. Such systems can learn and transfer desired speaking styles from one seen speaker to another (in multi-style multi-speaker settings), which is highly desirable for creating scalable and customizable Human-Computer Interaction systems. In this work we explore one-to-many style transfer from a dedicated single-speaker conversational corpus with style nuances and interjections. We elaborate on the corpus design and explore the feasibility of such style transfer when assisted with Voice-Conversion-based data augmentation. In a set of subjective listening experiments, this approach resulted in high-fidelity style transfer with no quality degradation. However, a certain voice persona shift was observed, requiring further improvements in voice conversion.    
\end{abstract}

\section{Introduction}
\footnotetext[1]{Authors are listed alphabetically and contributed equally.}

Text-to-speech (TTS) systems based on sequence-to-sequence (S2S) and neural vocoding techniques~\cite{Wang:2017:tacotron,Shen-Pang:18,Shechtman-Sorin:19, Shen-Jia:20} produce state-of-the-art performance that approach or match human levels, and have led to a ubiquitous deployment of the technology in speech interfaces. In tandem with the evolving quality of such systems, we see a move toward solving more complex cases that arise in Human-Computer Interaction (HCI), including the need to synthesize different styles for various usage contexts, and to transfer this ability across various  speakers and languages under often limited-data conditions. This work pursues several of these goals by creating fluid, conversational voices that can be naturally deployed within dialogue systems, are capable of a range of expressive devices observed in human-to-human communication (including the generation of non-lexical interjections), and can be learned from a conversational corpus and extended to novel voices lacking such training material. Central to the approach is the use of a style-preserving Voice Conversion (VC) algorithm that replicates a carefully-designed conversational TTS corpus in new target voices and enables transplantation of conversational characteristics (acoustic, lexical and non-lexical) across speakers. In order to allow this approach to be successful, this work also describes a strategy for collecting high-quality expressive data under controlled conditions that still respects the naturalness and spontaneity of live data.
 
Previous work in expressive S2S TTS has focused on varying notions of expressions, ranging from the implicit expression contained in a prosodic reference~\cite{SkerryRyan-Battenberg:18} to the use  of explicit, but fairly polar, styles that may be insufficient for dialogues~\cite{Shechtman-Fernandez:21b}. 
Learning conversational styles from spontaneous dialogues without dedicated TTS corpora was tackled in~\cite{Ben-David-Shechtman:21}, illustrating the difficulty and limitations of the approach, and justifying the work presented here using controlled data. The closest research to ours to explicitly target conversational TTS is the work of~\cite{Guo-Zhang:21} for voice agents in Mandarin. That work also importantly deals with the issue of designing appropriate training corpora though our approach differs substantially from theirs in the explicit use of Dialog Act tags and collection methodology (we cover more of those differences in \S~\ref{sec:corpus}). The synthesis of interjections 
finds precedent in earlier work by~\cite{Adell-Escudero:12} for unit-selection architectures, and more recently by~\cite{Szekely-Henter:19} for S2S frameworks. Our work differs in that it situates the generation within the conversational domain and targets a larger inventory. Another feature needed for effective synthesis of dialogue turns is the ability to realize narrow emphatic focus on demand, which we have previously presented in~\cite{Shechtman-Fernandez:21} and continue to use here. 

The idea of using VC as data-augmentation has been previously explored within tasks like the classification of symbolic prosodic events~\cite{Fernandez-Rosenberg:17}, recognition of low-resourced accented speech~\cite{Fukuda-Fernandez:18}, and more recently  in~\cite{Huybrechts-Merrit:21} to address a similar goal as in this work to mitigate the lack of expressive material for a variety of target speakers. Our work differs in the fact we tackle a more fine-grained set of conversational events with more expressive range, and in the way we allocate training resources (we use a larger corpus with a richer set of tags from a single training speaker vs.~smaller corpora from several speakers).

We cover the details of designing and collecting expressive dialogues in \S\ref{sec:corpus}. In \S\ref{sec:arch} we present a S2S architecture sensitive to style transplant, and cover various techniques to enable this in \S\ref{sec:xplant}. The experimental setup in \S\ref{sec:experiments} describes the training mechanism for building and selecting the models evaluated in \S\ref{sec:eval} before concluding with observations and future directions in \S\ref{sec:conclusions}.

\section{Design of a TTS Conversational Corpus}
\label{sec:corpus}

An objective in this work is to assemble a suitable corpus of recorded utterances to develop conversational TTS systems, particularly for conversational voice agents in customer-care (CC) settings. There are opposing constraints at work here. On one hand, we strive for a style that feels spontaneous, lively, and with more relaxed articulation. Corpora of spontaneous, natural dialogues with these characteristics already exist~\cite{SWB:97}. However, unconstrained conversations also exhibit behaviors that pose challenges when building TTS systems and severely degrade its quality, like disfluencies, repetitions,  hypo-articulation, self-corrections,  intonationally ``ill-posed''  utterances, etc. 
Our solution is to adopt the classical approach of recording professional voice talent under controlled 
conditions, but to carefully draft a script that facilitates and elicits a range of expressiveness and discourse-level behaviors, and to work with a voice actor (VA) that can convincingly bring the script to life in a style that matches the CC setting. To that end, we have designed a script with the following:

\noindent {\bf (1) Appropriate structure and contextual history:} The main part of the script consists of two-part dialogues with the VA playing the role of an \textit{Agent} interacting with a \textit{User}. The VA has access to both sets of turns, but records only their lines while silently reading over the \textit{User} responses. Giving the VA access to the history
allows them to follow a narrative arch and react expressively to the ever-evolving context. The split between \textit{Agent} and \textit{User} is approximately 75\% -25\%. Though this figure can differ substantially from real-life interactions, it is done to  maximize the amount of audio collected within a given budget. The \textit{User} turns merely provide context and keep the interaction moving along, creating a space for the VA to perform.

\noindent {\bf (2)  Appropriate semantic context:} The script has been sourced from real dyads (text-based and transcribed conversations) from proprietary materials reflecting a variety of CC topics, as well as from open-source dialogues from the Taskmaster-2 corpus~\cite{Byrne-Krishnamoorthi:19}. These have been edited as needed for semantic clarity, to eliminate any  personal  information, and in order to reduce the artifacts mentioned (repetitions, corrections, etc.). 

\noindent {\bf (3) Dialog-Act Tagging:} The script has been annotated with ten tags reflecting various communicative and dialog-managing functions. This is done both to exploit higher-level features during modeling (see \S~\ref{sec:arch}), but also to aid the VA (who has access to the marked-up text while recording) understand the structure of the dialogues and to guide them toward particular stylistic performances when enacting individual tags. Prior to recording the script, the VA and producer held training sessions to establish a common understanding of the meaning of these tags, and to workshop some distinct stylistic renderings for them.

\noindent {\bf (4) Interjections:} A substantial portion of the script contains a set of seven interjections (\S~\ref{ssec:das}) to mirror several exclamations, fillers, etc. observed in naturally occurring dialogues, and to inject convincing naturalness and spontaneity into the dialogues.

A strategy was proposed in~\cite{Guo-Zhang:21}  for collecting dialogues for TTS with similar goals of maximizing spontaneity while reducing performance artifacts. Both that approach and ours eschew using isolated turns in favor of contextual dialogues as an elicitation paradigm. However, \cite{Guo-Zhang:21} allows freer improvisation during the recording between two speakers whereas we allow the VA to deviate from the script only in minor ways (corrected afterwards), and provide guidance with DA tags to hit desired expressiveness targets. With this design 100\% of the recorded audio is usable to train an \textit{Agent}-like model whereas~\cite{Guo-Zhang:21} uses only the \textit{Agent} turns amounting to about 50\% of the collected data. For our work, we collected about 7 hours of audio; this includes 6 hours of conversations, and 1 hour of non-conversational, but expressive material for synthesizing focal emphasis. 

\subsection{Dialog Acts and Interjections}
\label{ssec:das}

In this work we propose a set of ten \textit{dialog tags} to help describe, control, 
and guide appropriate expressive realizations of the events happening in a dialogue. We approached the problem of creating such an inventory from both theoretical and empirical angles. The DIT++ taxonomy of Dialog Acts (DAs)~\cite{DIT}, which provides a multi-dimensional framework for analyzing the communicative functions of acts (and has a focus on both human-human (H2H) and human-machine (HM) interactions) was consulted to have an initial set of descriptors. This inventory is quite large and likely over-specialized for synthesis goals, so the next step was to cross-reference these labels against actual HM dialogues to empirically select or modify the tags to arrive at a more concise set. A fundamental selection principle was to focus on commonly occurring DAs that would require, or benefit from, a dedicated stylistic shaping in terms of both acoustic realization (i.e., prosody and voice quality) and linguistic content (i.e., formulaic and idiomatic constructions that can serve as affective carriers, and allow for the inclusion of interjections. This led to the following inventory: \{\textit{Agreement}, \textit{Farewell}, \textit{Greeting}, \textit{Empathy}, \textit{Instruction}, \textit{Positive Feedback}, \textit{Surprise}, \textit{Thanks}, \textit{Uncertainty}, \textit{Waiting} \}.

By interjections we mean a set of often non-lexical items that are used
as exclamations, fillers, and backchannels, and which are prevalent, and indeed
necessary, in conversational settings.  Interjections can
differ from standard lexical items in their segmental and suprasegmental characteristics; for instance: unusual durations or pitch excursions governed by pragmatic factors (and not fully predictable from syntax), little to no voicing during production, etc. 
Even when we limit the inventory to items that can be transcribed phonetically
fairly accurately (excluding more complex sounds like laughter), their high-quality
synthesis from a corpus without enough exemplars remains challenging because of the differences just mentioned. To enable their successful synthesis during a dialogue, we have included the following set of interjections in the design of the corpus, where they appear in combination with a variety of DA tags previously listed: \{\textit{aha},  \textit{oh},  \textit{hmm}, \textit{huh}, \textit{uh}, \textit{uh-huh}, \textit{um}\}. This inventory is by no means exhaustive and has been designed to attempt coverage of various desirable functional roles within a dialogue while trying to retain a manageable inventory\footnote{More details on the definitions and usage of the individual DA tags and interjections can be found in  \href{http://ibm.biz/IS22-S2SConv}{http://ibm.biz/IS22-S2SConv}.}. 

\section{Architecture Description}
\label{sec:arch}

\begin{figure}[t]
  \centering
  \includegraphics[width=\linewidth]{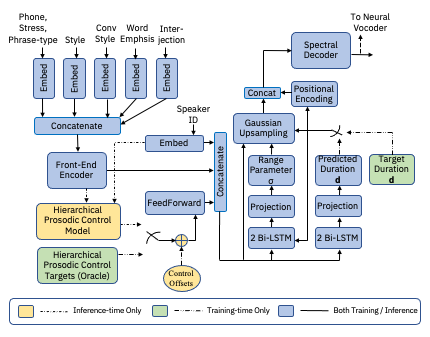}
  \setlength{\abovecaptionskip}{-12pt}
  \setlength{\belowcaptionskip}{-12pt}
  \caption{
  The proposed HPC-controllable non-attentive Tacotron2 acoustic model, predicting LPCNet acoustic features from the extended phone sequence.}
 \vspace{-2mm}
  \label{fig:arch}
\end{figure}

Our model (Fig.~\ref{fig:arch}) is a variant of the non-attentive Tacotron2~\cite{Shen-Pang:18} architecture (NAT2) from~\cite{Shen-Jia:20}
with an extended set of symbolic input embeddings,
augmented with hierarchical prosodic controls 
(HPCs)~\cite{Shechtman-Fernandez:21,Shechtman-Fernandez:21b} to improve its controllability. We previously showed that these HPCs are able to realize lexical focus~\cite{Shechtman-Fernandez:21}, and facilitate speaking-style transfer (for \textit{good news} and \textit{apology}) in a \textit{attentive} Tacotron2-like architecture~\cite{Shechtman-Fernandez:21b} coupled with an LPCNet~\cite{Valin-Skoglund:19} neural vocoder. In this work we combine NAT2
with the controllability provided by HPC~\cite{Shechtman-Fernandez:21,Shechtman-Fernandez:21b}. Large scale crowd MOS tests (up to 100 subjects, 40 samples per system, 25 votes per sample) revealed that the proposed HPC-controlled \textit{NAT2} architecture  produces speech with the same quality and naturalness as the HPC-controlled \textit{attentive} Tacotron2 architecture~\cite{Shechtman-Fernandez:21,Shechtman-Fernandez:21b} while almost entirely eliminating stuttering, repetitions, early stopping and other stability problems affecting the attentive Tacotron2 models in some edge cases. We also ensured that the proposed HPC-controlled NAT2 system is responsive to HPC offsets~\cite{Shechtman-Fernandez:21b} and suitable for local word focus realization, as in~\cite{Shechtman-Fernandez:21}, and for "simpler" speaking style transfer, as in~\cite{Shechtman-Fernandez:21b}, and is thus a fair baseline for the challenging conversational style-transplantation task.      

The proposed S2S model is fed with phonetic symbols including pauses and word boundaries, enriched with lexicographic stress, phrase type (as extracted from the input text by our proprietary TTS Front-End) and additional prosodic directives, that were exposed to some of the voice talents during some of the speech recording sessions (i.e. word emphasis~\cite{Shechtman-Fernandez:21}, speaking styles~\cite{Shechtman-Fernandez:21b} and unique for this work \textit{dialog tags} and \textit{interjection types}) and generates an acoustic feature vector sequence that is then fed to an independently trained 
LPCNet~\cite{Valin-Skoglund:19} 
neural vocoder to generate high quality samples in real time~\cite{Shechtman-Rabinovitz:20}. The sequence to sequence acoustic model (see Fig.~\ref{fig:arch}) is first trained with ground truth HPCs,  minimizing the acoustic regression loss, as in~\cite{Shechtman-Fernandez:21b}, but based on $L_1 + L_2$ loss operator~\cite{Shen-Pang:18}, plus the duration prediction loss~\cite{Shen-Jia:20}, then the HPC prediction sub-model is trained~\cite{Shechtman-Fernandez:21b}.


The phone identity, lexical stress and phrase type are represented by a single embedding, while the speaking style (e.g. \textit{neutral} or \textit{conversational}), the word emphasis~\cite{Shechtman-Fernandez:21}, the dialog tags and the interjection types (corresponding, respectively, to the DA and interjection inventories from \S~\ref{ssec:das}) each has its own embedding (see Fig.~\ref{fig:arch}). 
For brevity we will refer to the concatenation of the front-end (FE) encoder outputs, the speaker-id embedding, and the HPC embedding as the \emph{encoder output}. 

The \textit{encoder output} is fed to the phone-level Duration Predictor and the Range predictor that facilitates the Gaussian upsampling of the \textit{encoder-output} sequence to match the length of the spectral output (as determined by ground-truth phone durations)~\cite{Shen-Jia:20}. The upsampled encoder output is concatenated with the positional encoder output and fed to the Tacotron2-like autoregressive LSTM decoder~\cite{Shen-Jia:20}, containing two LSTM stacks, followed by two parallel Linear layers predicting the 80-dim. mel-cepstral features (serving for the autoregressive decoder feedback) and the 22-dim. LPCNet features (20 for spectral envelope and 2 for pitch). The predicted LPCNet features are further processed with two post-nets 
(one for the cepstrum, and one for the pitch parameters) and are fed to the separately trained LPCNet neural vocoder, as detailed in~\cite{Shechtman-Rabinovitz:20}.    



\section{Conversational Style Transplantation}
\label{sec:xplant}

In this work we aim to transplant the conversational style with interjections from a single speaker to others. This is challenging due to asymmetric data conditions, as the conversational-style and speaker-ID labels are highly correlated (over 80\% of the conversational-speaker data is of \textit{conversational} style) and also associated with specific textual patterns, such as common phrases and interjections that only arise in the conversational corpus. Thus the plain training of the proposed HPC-controlled NAT2 system alone (\S\ref{sec:arch}) does not attain acceptable style transfer to new speakers (\S\ref{sec:eval}) and leads to occasional persona switching for some inputs. To decouple the speaker ID from the rest of FE inputs and labels we explored the following Speaker Adversarial training and VC-based data-augmentation approaches.          

\subsection{Speaker Adversarial Classifier}

To prevent speaker ID leakage into the speaker-independent FE Encoder, we deploy a phone-level adversarial speaker classifier~\cite{Nekvinda-Duvsek:20} with inverse gradient layer~\cite{ganin2016domain} applied to the encoder outputs during training. The cross-entropy of speaker predictions is added to the total training loss as an additional optimization task. However, while improving the speaker similarity, this method significantly deteriorates the style similarity (see \S\ref{sec:eval}).     


\subsection{Data Augmentation by Voice Conversion}

The undesired coupling between the conversational style and the conversational speaker ID can be implicitly reduced by training with more multi-speaker conversational data. In~\cite{Huybrechts-Merrit:21} VC was proposed as a data-augmentation technique to increase the existing amount of multi-speaker data with a desired style. The trained multi-speaker multi-style TTS system is further fine-tuned into a single-style single-speaker system with a small amount (0.5 hours) of the target style and target speaker recordings available~\cite{Huybrechts-Merrit:21}. In the current work, however, no target-speaker recordings with target-style rendering exist, so we explore style transplantation solely relying on VC-generated data: we  apply VC on the conversational style corpus to augment the training set with synthetic conversational style utterances for all the existing non-conversational speakers, and remove the original conversational voice recordings from the training completely. To that end, we developed a prosody-preserving VC system, similar to~\cite{Karlapati-Moinet:20}, but based on the pretrained HuBERT~\cite{hsu2021hubert} speech representation, thus not requiring the phonetic sequence as an input. The proposed VC system is tailored to the S2S TTS architecture, has a simpler training procedure than in~\cite{polyak2021speech}, and exploits an alternative HuBERT feature discretization technique to facilitate Speaker ID disentanglement.  



Let $X = [x_1,\cdots, x_T ]$ denote a speech utterance of $T$ frames and let $S$ denote its speaker identity.
Our VC system is a conditional Discrete Variational Autoencoder. During training, $s$ corresponds to the actual speaker of $X$ whereas at inference time, the conversion is achieved by forcing $s$ to be a different speaker. The model is composed of the following parts:
\begin{itemize}[leftmargin=*]
    \item Encoder: The encoder is a pretrained HuBERT-Base model~\cite{hsu2021hubert}.
    The first 6 layers are fixed during training while the rest 6 layers are fine-tuned. In addition, Gaussian upsampling~\cite{Shen-Jia:20} is used to resample from HuBERT frame rate to the TTS frame rate, thus preserving the original phone alignment. 
    
    \item Discretization: In order to produce more stable representation, the encoder output is mapped from 768 into $N\times 2$ space, and discretized into $N=32$ binary variables. To keep the discrete variable differentiable, we used the Gumbel-Softmax relaxation \cite{maddison2016concrete, jang2016categorical} with a fixed temperature of 1. Let $h(x_t)$ denote the discrete outcome of the frame $x_t$
    
    \item Speaker integration: The discrete representation is forwarded to the decoder conditioned on the speaker identity $s$. We represent $s$ as a one-hot vector and compute the latent $z_t = s_t \otimes h(x_t)$ where $\otimes$ is the outer product operation. We found this integration technique more effective than concatenation.

    \item Spectral decoder: The architecture for the decoder and the reconstruction loss function are the same as used in the proposed TTS model (\S\ref{sec:arch}), generating the acoustic features for the augmented training data.

\end{itemize}

\begin{figure}[t]
  \centering
  \includegraphics[width=\linewidth]{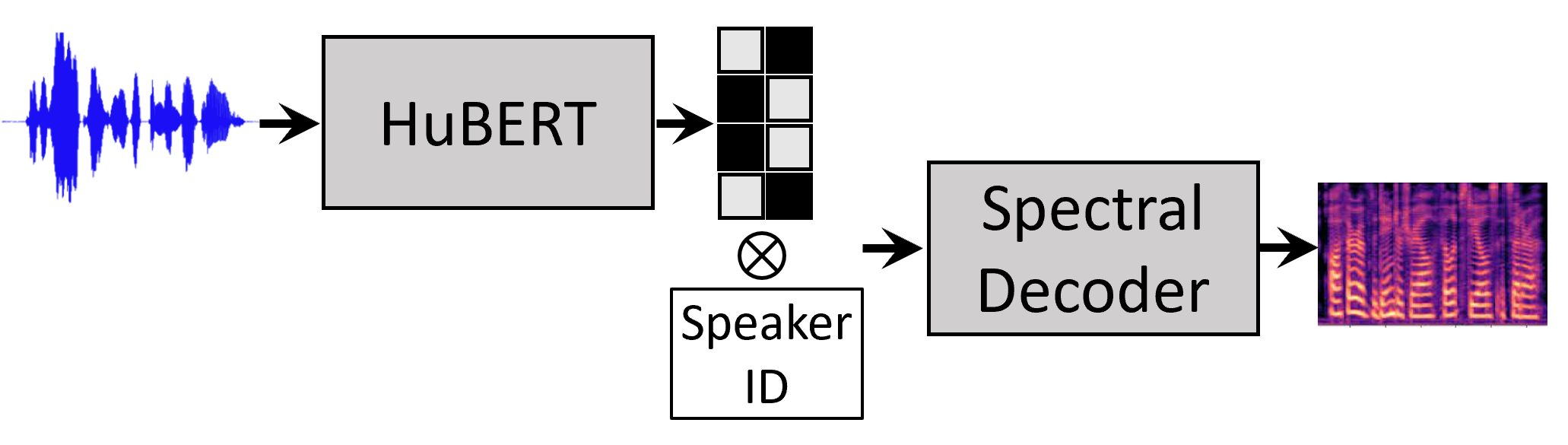}
  \setlength{\abovecaptionskip}{-10pt}
  \setlength{\belowcaptionskip}{-10pt}
  \caption{Voice Conversion System for \textbf{Aug} TTS}
  \label{fig:arch2}
\end{figure}

\section{Experimental Setup}
\label{sec:experiments}

The training material comprises wide-band (22.05 kHz) speech corpora from three professional native speakers of US English uttered mostly in neutral speaking style in various domains, two females and one male, of 10-17k sentences each, and the conversational US English female voice corpus of 4k sentences (\S\ref{sec:corpus}). Both of the non-conversational female voices were selected to evaluate conversational speaking-style transplantation. We trained the following multi-speaker TTS systems sharing the described architecture to assess different transplantation techniques. All systems were fine-tuned by adjusting the constant Prosody Control Offsets (as described in \cite{Shechtman-Fernandez:21b}). 
\begin{itemize}[leftmargin=*]
\item {\bf Base}: A TTS system trained just on the recorded data.
\item {\bf Adv}: The baseline TTS system with Adversarial Speaker Classifier~\cite{Nekvinda-Duvsek:20}, applied on the Front-End Encoder output (trained just on the recorded data).
\item {\bf Aug}: The baseline TTS system, trained on the recorded voice corpora, excluding the conversational voice corpus, and augmented with the VC-generated conversational data (for all the three neutral speakers in the training corpora).  
\end{itemize}

\subsection{Automated Checkpoint Selection}

Due to training-synthesis conditions mismatch in autoregressive architectures, selecting a final configuration and checkpoint can be time-consuming and require subjective listening~\cite{Shechtman-Haws:21}. To facilitate this, we introduce an automated-selection technique applied to each system. The central element of this technique is an objective assessment of a model checkpoint based on synthesizing speech from the texts corresponding to a held-out set of audio recordings taken as reference. For each synthesized sentence a correct falling/rising intonation pattern is assessed. To this end, for each synthesized-and-reference pair, the pitch contour slope is estimated within the time interval occupied by the last two words. The slope is estimated using a weighted linear regression of the $f_0$ contour with the weights set to the voicing values. Opposite signs of TTS and reference slope values suggest unnatural intonation patterns. All checkpoints, possibly corresponding to different model configurations, are ranked based on the counts of unnatural patterns found in all synthesized utterances. One or few top-ranked checkpoints are then selected for validation by human listeners.

\section{Evaluation}
\label{sec:eval}

We designed a set of subjective evaluations to assess how well the conversational style is transferred within various multi-speaker TTS systems using a test script of 70 conversational utterances simulating real CC use cases (for a speaker-similarity test only half were used), and manually annotated with interjections, dialog tags and word emphasis labels\footnote{Audio samples along with the corresponding annotated text inputs are available at \href{http://ibm.biz/IS22-S2SConv}{http://ibm.biz/IS22-S2SConv}.}.  

\begin{table}[h]
  \setlength{\belowcaptionskip}{-8pt}
  \setlength{\abovecaptionskip}{2pt}
  \caption{Speaking style similarity ABX (Preference scores [\%]) }
  \label{tab:style_test}
  \centering
  \begin{tabular}{cccccc} \toprule
Speaker & Sys1/Sys2 & Sys1 & No Pref. & Sys2 & \textit{p}-value \\ \midrule
  \multirow{3}{*}{F1} & \textbf{Aug}/\textbf{Base} & \textbf{61.11} & 10.51 & 28.39 & $<$0.01 \\ \cmidrule{2-6}
                      & \textbf{Aug}/\textbf{Adv} &\textbf{ 63.33} & 8.93 & 27.74 & $<$0.01 \\ \cmidrule{2-6}
                      & \textbf{Adv}\textbf{/Base}  & 43.10 & 11.83 & 45.08 & 0.232 \\ \midrule
   \multirow{3}{*}{F2} & \textbf{Aug}/\textbf{Base} & \textbf{54.68} & 12.99 & 32.34 & $<$0.01 \\ \cmidrule{2-6}
                      & \textbf{Aug}/\textbf{Adv} & \textbf{60.09} & 12.05 & 27.86 & $<$0.01 \\ \cmidrule{2-6}
                      & \textbf{Adv}/\textbf{Base}  & 36.65 & 15.86 & \textbf{47.49} & $<$0.01 \\
    \bottomrule
  \end{tabular}
\vspace{-6mm}
\end{table}

All the tests in this work are designed to be parallel, so corresponding high quality synthetic TTS outcomes serve as test references. Specifically, we assess overall quality and naturalness by 5-scale MOS tests with neutral synthetic references (generated by {\bf Adv}); speaker similarity by 5-scale ([-2:2]) MOS similarity test with two speaker references: \textit{A}-reference (conversational speaker's synthetic samples, generated by {\bf Base}) and \textit{B}-reference (target speaker's neutral synthetic samples, generated by {\bf Adv}), where subjects were presented with four scale categories from ~\cite{wester2016analysis} plus \textit{neither-A-nor-B} category in the middle (zero-value); and speaking style similarity by ABX tests with style reference (synthesized by {\bf Base} with the conversational speaker ID). In the latter case, subjects were shown groups of three parallel  audio  samples, containing a pair of test samples generated with a target speaker ID and a reference-style sample generated with the conversational-speaker ID. (The reference-style samples were found to convincingly express the conversational speaking style, scoring 4.4 in a separate MOS test). All tests were conducted in crowd-sourced settings where subjects were screened with a transcription test to filter out insufficient English proficiency and/or bad listening conditions. Eventually, about 1000 votes per system were used in the style-similarity and MOS-quality tests, and about 500 votes per system were used in the speaker-similarity tests.   

\begin{table}[h]
  \setlength{\belowcaptionskip}{-8pt}
  \setlength{\abovecaptionskip}{2pt}
  \caption{Speaker similarity [-2:2] with 95\% conf. interval} 
  \label{tab:spk}
  \centering
  \begin{tabular}{cccc} \toprule
   Spk/Sys & \textbf{Base} & \textbf{Adv} & \textbf{Aug} \\ \midrule
    F1 & 0.40 ± 0.11 &\textbf{ 1.35} ± 0.09 & 0.42 ± 0.11   \\ \midrule
    F2 & 0.92 ± 0.11 & \textbf{1.41} ± 0.09 & 0.82 ± 0.11   \\ 
    \bottomrule
  \end{tabular}
\vspace{-2mm}
\end{table}



\begin{table}[h]
  \setlength{\belowcaptionskip}{-5pt}
  \setlength{\abovecaptionskip}{2pt}
  \caption{Quality \& naturalness MOS with 95\% conf. interval} 
  \label{tab:mos}
  \centering
  \begin{tabular}{ccccc} \toprule
    Spk/Sys & Neutral & \textbf{Base} & \textbf{Adv} & \textbf{Aug} \\ \midrule
    F1 & 3.77±0.05 & 3.58±0.06 & 3.79±0.05 & \textbf{3.96}±0.05 \\  \midrule
    F2 & 3.91±0.05 & 3.77±0.05 & 3.89±0.05 & 3.94±0.05 \\  
    \bottomrule
  \end{tabular}
\vspace{-6mm}
\end{table}

\section{Conclusions}
\label{sec:conclusions}
The style similarity (Table~\ref{tab:style_test}) and quality (Table~\ref{tab:mos}) evaluations reveal that \textbf{Aug} system significantly outperforms all the other systems in terms of transplanted style similarity while not degrading the overall naturalness. However, a speaker identity shift in \textbf{Aug} is still perceivable (Table~\ref{tab:spk}). It is clear, though, that further improvement in the VC system quality would directly contribute to the resulting TTS speaker similarity improvement. 
This defines the main direction of our ongoing and future research, which also includes addressing the more difficult case of female-to-male style transplantation. We have also assumed in our evaluation a test case where user-input mark-up drives the different conversational sub-style renderings. A natural future extension to this work is to do this automatically to either suggest sensible defaults or to obviate this step from the user.

\bibliographystyle{IEEEtran}

\bibliography{vc_aug}
\end{document}